\documentclass[aps,pra,twocolumn,superscriptaddress,groupedaddress]{revtex4} 

\usepackage{graphicx}  
\usepackage{float}
\usepackage{dcolumn}   
\usepackage{bm}        
\usepackage{amssymb}   
\usepackage{amsmath}
\usepackage{isomath}
\usepackage[usenames, dvipsnames]{color}
\usepackage{xcolor}
\usepackage{soul}
\usepackage{physics}
\usepackage[normalem]{ulem}
\usepackage{balance}
\usepackage{hyperref}
\hypersetup{
     colorlinks   = false,
     citecolor    = blue
}
\usepackage{lipsum}
\usepackage{amsmath,esint}
\newcommand{\D}{\mathrm{\Delta}}
\DeclareMathOperator{\sgn}{sgn}

\begin{document}

\title{Elastic Weyl points and surface arc states in 3D structures}

\author{Xiaotian Shi}
\affiliation{Aeronautics and Astronautics, University of Washington, Seattle, WA 98195, USA}
\author{Rajesh Chaunsali}
\affiliation{Aeronautics and Astronautics, University of Washington, Seattle, WA 98195, USA}
\affiliation{LAUM, CNRS, Le Mans Universit\'{e}, Avenue Olivier Messiaen, 72085 Le Mans, France}
\author{Feng Li}
\affiliation{South China University of Technology, School of Physics and Optoelectronic Technology, Guangzhou, Guangdong, 5140640, China}
\author{Jinkyu Yang}
\thanks{jkyang@aa.washington.edu}
\affiliation{Aeronautics and Astronautics, University of Washington, Seattle, WA 98195, USA}


\begin{abstract}
The study of Weyl points in electronic systems has inspired many recent researches in classical systems such as photonic and acoustic lattices. Here we show how the Weyl physics can also inspire the design of novel elastic structures. We construct a single-phase 3D structure, an analogue of the AA-stacked honeycomb lattice, and predict the existence of Weyl points with opposite topological charges (${\pm 1}$), elastic Fermi arcs, and the associated gapless topologically protected surface states. We employ full-scale numerical simulations on the elastic 3D structure, and present a clear visualization of topological surface states that are directional and robust. Such designed lattices can pave the way for novel vibration control and energy harvesting on structures that are ubiquitous in many engineering applications. 
\end{abstract}
\flushbottom
\maketitle
\thispagestyle{empty}


\section*{INTRODUCTION}
Phononic crystals and metamaterials have shown new and exciting ways to control the flow of wave propagation in the medium~\cite{1,2,3,4}. Recently, the topology of band structures has emerged as a new design tool in this context. The essential idea is to characterize the bulk dispersion topologically and predict its implications on the edges/surfaces of the system. A nonzero topological invariant of the bulk usually implies the existence of edge or surface waves with nontrivial properties, such as directionality and robustness~\cite{5,6,7}. Several interesting strategies to manipulate elastic waves have thus been shown \cite{8,9}. However, the studies so far focused mainly on 1D and 2D systems. It is not clear how a 3D elastic structure could be designed to support topological surface states. What special characteristics those surface state would have and how they could be harnessed in engineering settings are some key questions. In this study, we attempt to address these questions by taking inspiration from the Weyl physics.

Weyl semimetals {\cite{10,11,12,13,14}} have recently attracted significant attention for their exotic features. In Weyl semimetals, the Weyl point refers to the degeneracy point of two bands having linear dispersion in all directions in the 3D reciprocal space. The effective Weyl Hamiltonian is, in general, given by $H(\textbf{k})=f(\textbf{k})\sigma_0+v_xk_x\sigma_x+v_yk_y\sigma_y+v_zk_z\sigma_z $, where $f(\textbf{k})$ is an arbitrary real function, $v_i$, $k_i$, and $\sigma_i$ represent group velocity, momentum, and Pauli matrix, respectively.  Weyl points behave as the sources or the sinks of the Berry curvature in the reciprocal space. By integrating the Berry flux on a sphere surrounding a Weyl point, we can get the non-vanishing topological charge (or Chern number) associated with it {\cite{15}}. The Weyl point is robust against small perturbations and cannot be easily gapped unless it is annihilated with another Weyl point with the opposite topological charge {\cite{16}}. For electronic systems, previous researches have shown many unusual phenomena associated with Weyl points, such as robust surface states \cite{10} and chiral anomaly \cite{17}. Later on, the Weyl physics has shown to be useful in the classical systems of photonic {\cite{16,18,19,20}} and acoustic lattices {\cite{21,22,23,24,25,26,27}}. 

The implementation of the Weyl physics in elastic structures, however, has been challenging so far. Recently, a self-assembled double gyroid structure that contains Weyl points for both electromagnetic and elastic waves was proposed~\cite{28}. Later, a design consisting of a thin plate and beams, which carries both Weyl and double-Weyl points, was also proposed~\cite{29}. 
In spite of these, the experimental demonstrations of elastic Weyl points remain elusive to date. Furthermore, there has not been even a study reporting full-scale numerical simulations in the elastic setting, by which the Weyl physics can be directly visualized and appreciated.
 This is due to the fact that such structures are extremely intricate to fabricate. At the same time, it is very demanding to computationally simulate their full-scale wave dynamics, because it involves with several types of elastic modes.

In this research, we design a 3D elastic lattice made entirely of beams, which allow both translational and rotational degrees of freedom along their length directions. We employ the finite element analysis (FEA) to obtain a dispersion diagram for the unit cell and discuss its topological features in relation to the Weyl physics. Inspired by widely used 3D hollow structures in engineering, e.g., fuselage, we construct a full-scale hollow structure and show the existence of topological surface states in it. We also elucidate the relation of their directionality with the elastic Fermi arcs in the reciprocal space. We perform a transient simulation of the structure to numerically demonstrate the propagation of nontrivial surface waves. This study therefore paves the way for future research on the design and fabrication of Weyl physics-based structures for engineering applications, such as vibration control and energy harvesting. 

\begin{figure}[t]
\centering
\includegraphics[width=3.4in]{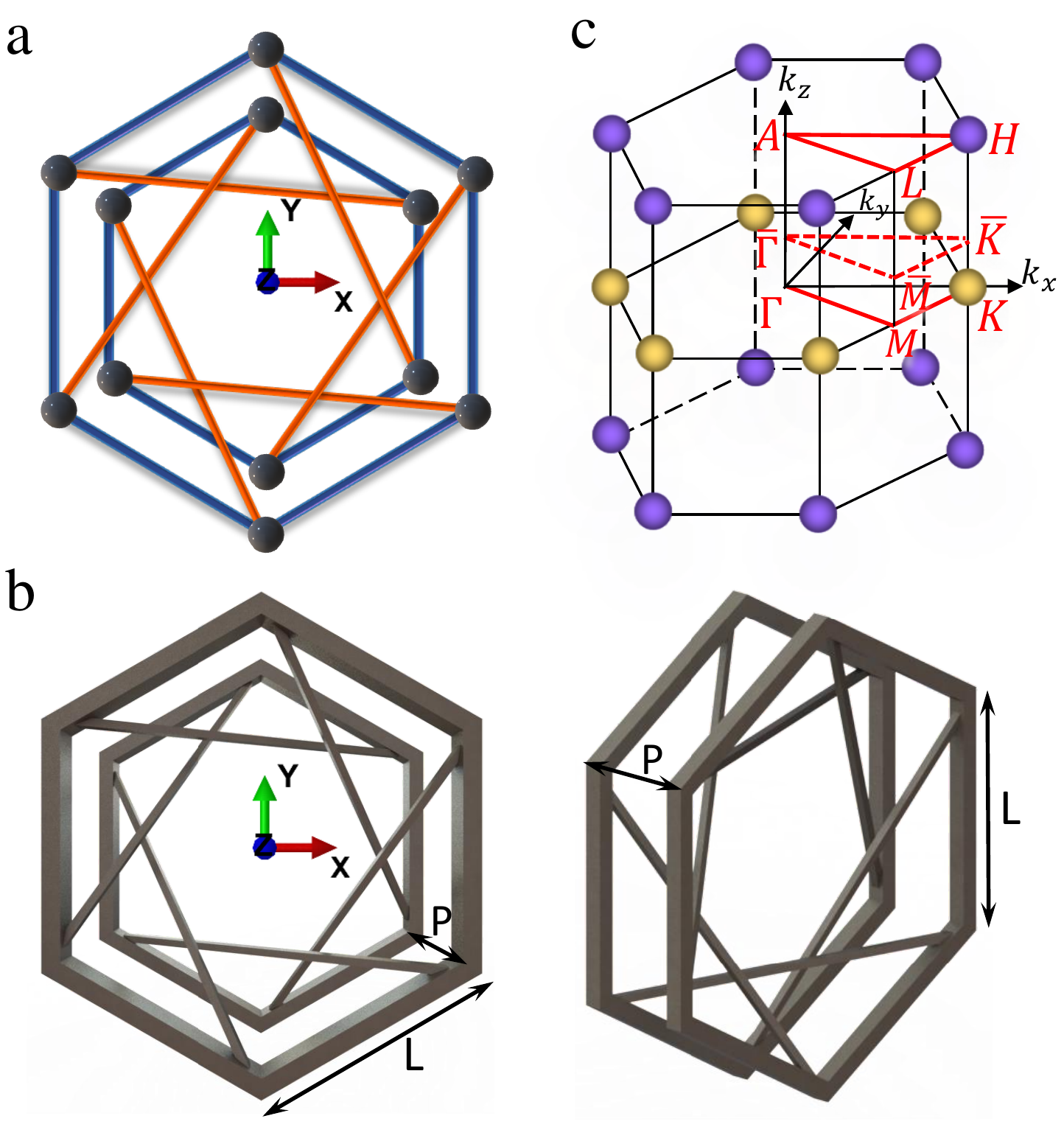}
\caption{[Color online] (a) AA-stacked hexagonal lattice (blue) with chiral inter-layer hopping (orange). (b) Top (left panel) and slanted (right) view of the unit cell of the 3D elastic structure. (c) Illustration of the first Brillouin zone and Weyl points with opposite topological charges indicated by the yellow and purple spheres.}
\label{fig:Figure1}
\end{figure}

\section*{DESIGN OF WEYL STRUCTURE} 
Previous studies in acoustics demonstrated the existence of Weyl points in a AA-stacked honeycomb lattice with chiral interlayer hopping \cite{21}. 
A schematic of its nearest-neighbor tight-binding model is illustrated in Fig. \ref{fig:Figure1}(a). 
To make an equivalent mechanical system, we would need to use masses and springs that are connected by hinge joints. However, for the more realistic design, we deviate from spring-mass description and propose a unit cell made of space beams as shown in Fig. \ref{fig:Figure1}(b).
We take beam length ${L = 20}$ mm and height $P= 10$ mm. All in-plane beams (parallel to $xy$ plane) have square cross section of width $3.0$ mm, while out-of-plane beams have square cross section of width $0.7$ mm to reduce the inter-layer stiffness. 
Note that we can still calculate the effective tight-binding Hamiltonian to analyze topological properties of our elastic structure (Appendix A). 
In Fig. \ref{fig:Figure1}(c), we show the first Brillouin zone with marked Weyl points at the high symmetric points. These are of two opposite charges (Appendix A) and will appear in the simulation results shown in the next section. 

To conduct the numerical simulation, we use the commercial FEA software ABAQUS. We model the space beams using the Timoshenko beam elements. 
We follow the method used in \cite{30} to apply periodic boundary conditions and calculate frequency band diagrams.
We use stainless steel 316L as the structural material with elastic modulus $E=180$ GPa, density $\rho=7900$ kg/m$^3$, and  Poisson's ratio $\nu=0.3$, which could be used for current 3D metal printing {\cite{31}}. We neglect any material dissipation.
The out-of-plane beams produce an effective synthetic gauge flux and break the effective time-reversal symmetry at a fixed $k_z$. Therefore, the system can be treated as an elastic realization of the topological Haldane model {\cite{32}.}

\begin{figure}[t]
\centering
\includegraphics[width=3.4in]{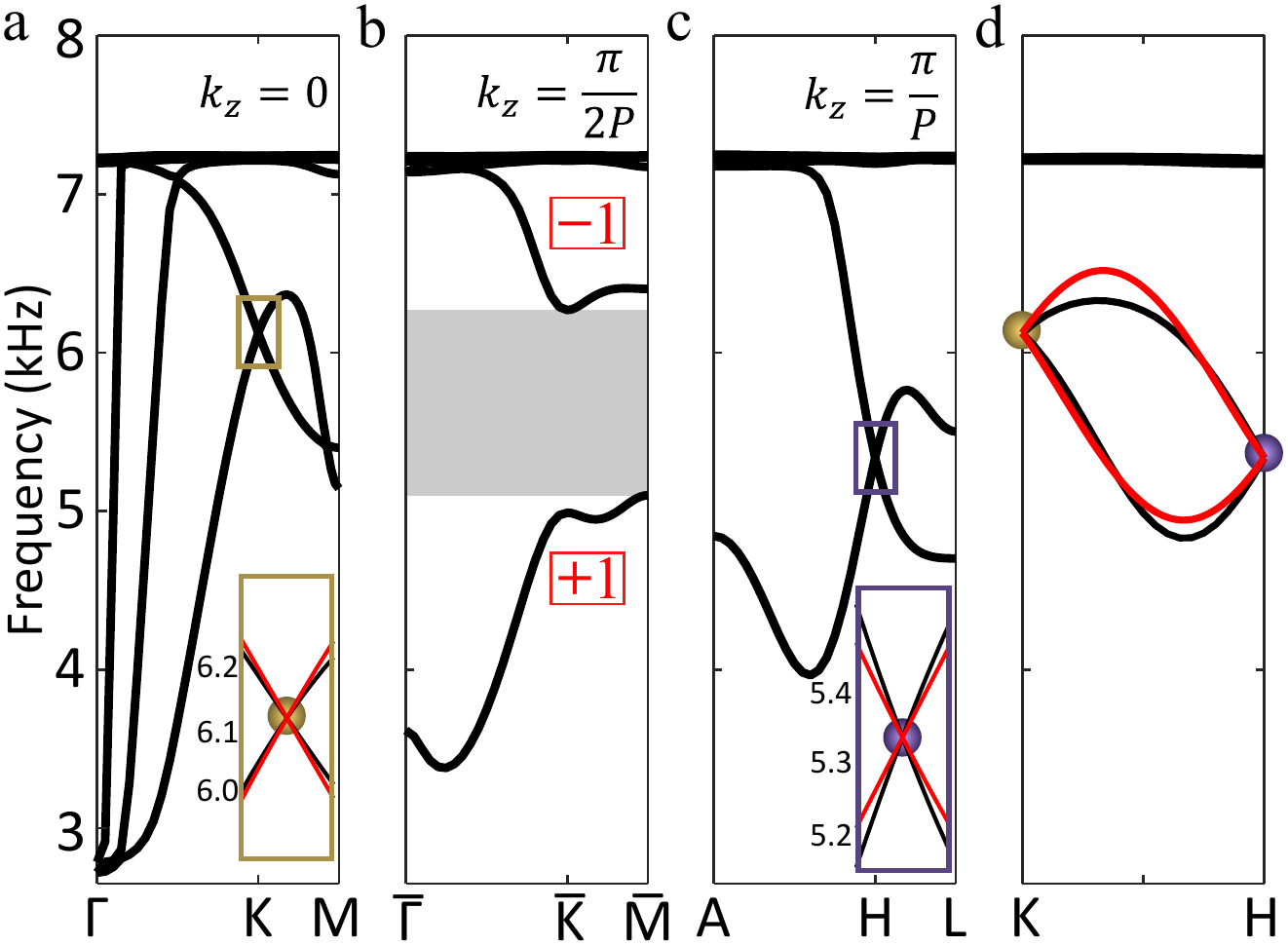}
\caption{[Color online] Dispersion diagrams on the reduced reciprocal $k_xk_y$ plane with fixed (a) $k_z=0$, (b) $k_z=\pi/(2P)$, and (c) $k_z=\pi/P$. The yellow (purple) sphere refers to the Weyl point located at the $K$ ($H$) point with topological charge $-1$ ($+1$). The grey area in (b) represents a complete band gap. (d) Dispersion diagram along the $KH$ line. The red and black curves are obtained from the two-band Hamiltonian and the FEA simulations, respectively. }
\label{fig:Figure2}
\end{figure}

\begin{figure*}[t]
\centering
\includegraphics[width=6.8in]{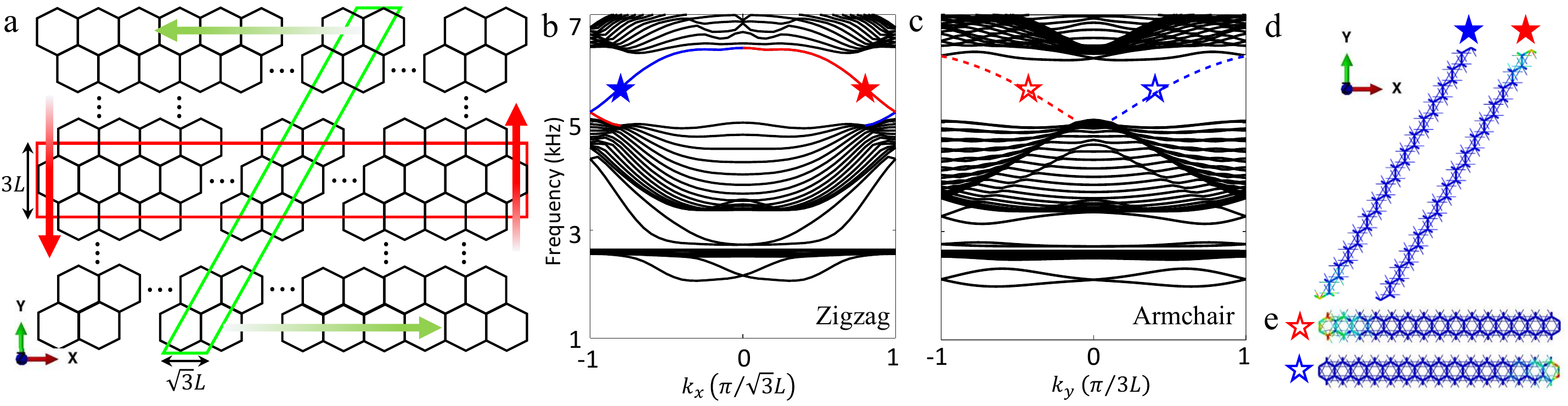}
\caption{[Color online] (a) Schematic of supercells with zigzag (green box) and armchair (red box) type boundaries. (b)-(c) Band structure with fixed $k_z=\pi/(2P)$ for the zigzag and armchair type supercells, respectively. The red and blue curves represent the topological surface modes at the two opposite ends of the supercell. (d)-(e) Mode shapes of the surface modes at 5.4 kHz corresponding to the solid and hollow stars in (b) and (c). Color intensity represents the magnitude of total displacements.}
\label{fig:Figure3}
\end{figure*}

\section*{RESULT}
\subsection*{Weyl points in unit-cell dispersion}
We show the band structure of the unit cell along the irreducible Brillouin zone at $k_z=0$ in Fig. \ref{fig:Figure2}(a).
We observe that the $13$th and the $14$th bands, predominately with out-of-plane polarization (see Appendix B), are degenerate at the $K$ point around $6.12$ kHz [see the rectangular box and its close-up inset in Fig. \ref{fig:Figure2}(a)]. 
This is a Weyl point in the system, and it is the same as the yellow spheres in Fig. \ref{fig:Figure1}(c). We calculate the Weyl charge by fitting the dispersion diagram of a two-band Hamiltonian around the Weyl point (see the red curves in the inset). As a result, we obtain the Weyl charge of $-1$ from this Weyl point (see Appendix A). 

When we plot the dispersion diagram for $k_z=\pi/(2P)$, the degeneracy of the bands is lifted, and there emerges a band gap between  $13$th and $14$th bands, as shown by the grey region in Fig. \ref{fig:Figure2}(b). We increase 	$k_z$ further to plot the dispersion curves at $k_z=\pi/P$ [Fig. \ref{fig:Figure2}(c)]. We observe that the band gap closes again, and the two bands establish a degeneracy at 5.34 kHz, but at the $H$ point of the Brillouin zone instead of the $K$ point. This is the second Weyl point in the system, and it corresponds to the purple markers  in Fig. \ref{fig:Figure1}(c). As shown in the inset, we again use the two-band Hamiltonian to represent the dispersion characteristics around this Weyl point and find that the Weyl point has $+1$ topological charge. 

In Fig. \ref{fig:Figure1}(d), we plot the dispersion diagram along the $KH$ direction to clearly visualize the Weyl degeneracy at the $K$ and $H$ points and the existence of a band gap between the $13$th and the $14$th bands when the $k_z$ value lies somewhere in between.
We note that the two-band Hamiltonian (red curves) captures this evolution of the bands obtained through the FEA simulations (black curves) reasonably well. We use this effective Hamitonian to numerically calculate the Chern numbers of the two bands above and below the band gap for a fixed $k_z$. For positive $k_z$, these are $-1$ and $+1$ for the upper and the lower bands, respectively, as marked in Fig.~\ref{fig:Figure2}(b). This indicates that the band gap is topologically nontrivial.

%

\subsection*{Directional surface states in supercell}
Based on the bulk-edge correspondence of topology, we expect topologically protected boundary modes arising at finite boundaries. 
To this end, we construct two types of supercells, consisting of 15 unit cells each, having both armchair and zigzag types of finite boundaries. 	
Figure \ref{fig:Figure3}(a) shows a schematic of how we choose the two types of the supercells. For the zigzag supercell (see the slanted green box), we apply periodic boundary conditions in both $x$- and $z$-directions. We use free boundary conditions at the top and bottom ends. We fix $k_z=\pi/2P$ and plot dispersion diagram in Fig. \ref{fig:Figure3}(b). We observe two modes inside the band gap, which are localized on the top (red) and the bottom (blue) ends of the supercell [see Fig. \ref{fig:Figure3}(d)]. Based on their slope we can conclude that the top (bottom) end mode would have a negative (positive) group velocity in the $x$-direction [see the green arrows in Fig. \ref{fig:Figure3}(a)]. 

Similarly, we study another supercell with the armchair type of boundaries [see the horizontal red box in Fig. \ref{fig:Figure3}(a)]. We show that it supports localized modes on the left and the right ends [see Figs. \ref{fig:Figure3}(c) and \ref{fig:Figure3}(e)]. These left and right end modes exhibit negative and positive group velocities, respectively [see the red arrows in Fig. \ref{fig:Figure3}(a)]. Therefore, it is straightforward to deduce that a wave packet injected at 5.4 kHz (shown as star) on the surface of the full-scale lattice, having simultaneous zigzag and armchair boundaries, would travel counterclockwise for $k_z=\pi/(2P)$. In the same vein, we expect to obtain a traveling surface wave in the clockwise direction for $k_z=-\pi/(2P)$.

\subsection*{Elastic Fermi arcs in full-scale model}
\begin{figure}[t]
\centering
\includegraphics[width=3.4in]{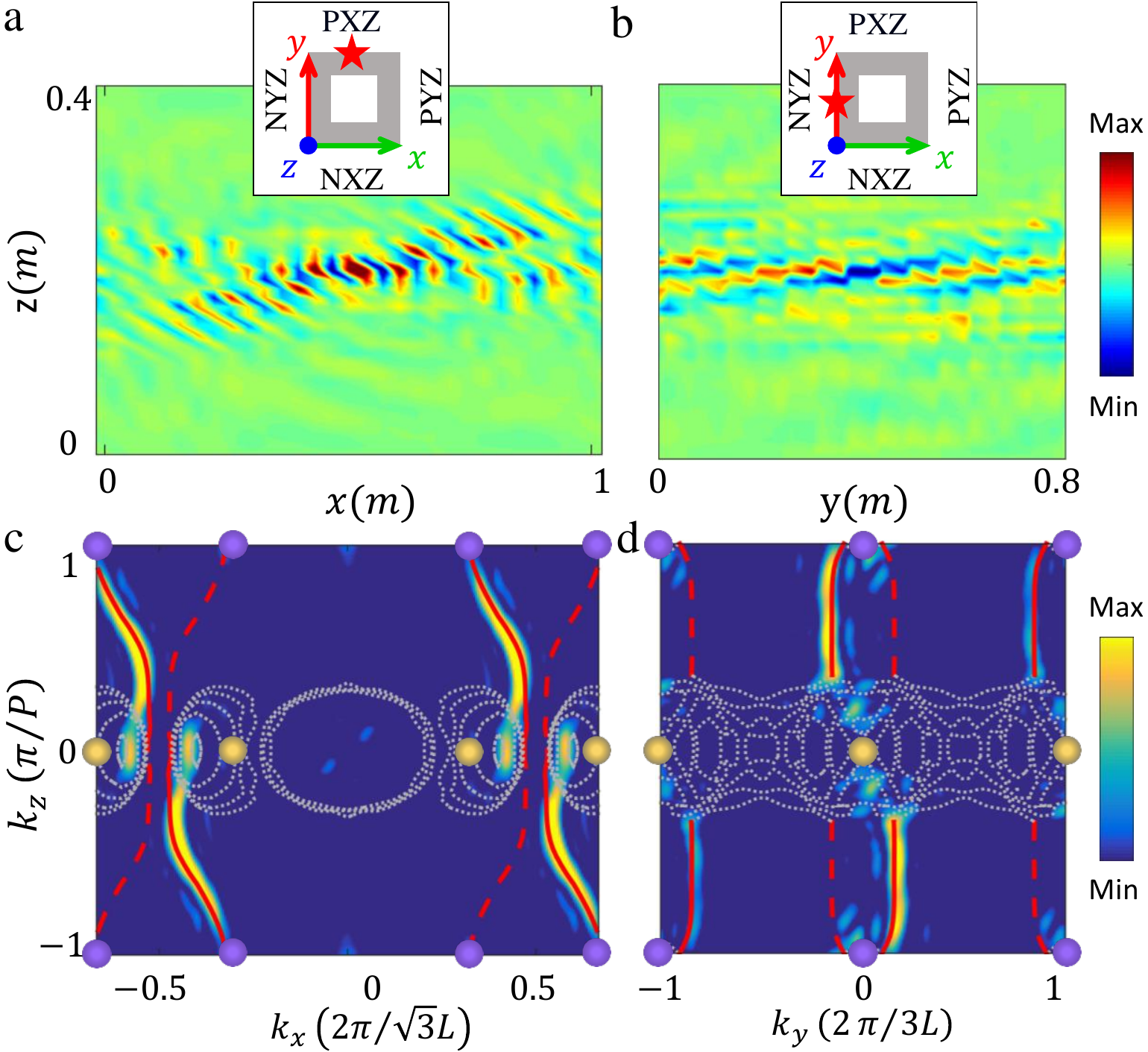}
\caption{[Color online] Surface states and elastic Fermi arcs in a full-scale 3D hollow structure. (a)-(b) Surface states under the harmonic excitation at $5.4$ kHz on the $xz$ and $yz$ planes, respectively. The point sources of excitation are placed in the center of each plane, as shown in the insets. The color intensity represents the nodal displacement  in the $z$-direction ($u_z$). (c)-(d) Spatial Fourier transforms of the field distributions of the surface states on the $xz$ and $yz$ planes, respectively. Spectral density shows the elastic Fermi arcs that connect the projections of the Weyl points with the opposite topological charges in the reduced 2D Brillouin zone. The red solid curves represent the simulated elastic Fermi arcs on the corresponding surfaces through supercell analysis, while the dashed curves indicate the Fermi arcs on the opposite surfaces. The projected bulk bands are shown as the dotted grey curves.}
\label{fig:Figure4}
\end{figure}

We now demonstrate the existence of surface states in a full-scale 3D structure. We choose a hollow structure for two reasons: (1) such structures are ubiquitous in applications; and (2) they require a reduced amount of computational time for numerical simulations, compared to solid ones. 
Without fixing $k_z$, we first excite our system at $5.4$ kHz in the $z$-direction. In Figs.~\ref{fig:Figure4}(a) and \ref{fig:Figure4}(b), we show the $z$-component of steady-state wave displacement when the excitation is placed at the centers of positive $xz$ (denoted as PXZ, see the red star mark in the inset) and negative $yz$ (NYZ) planes, respectively.  As we can see, the surface states propagate in particular directions and do not spread all across the whole plane. Especially in Fig.~\ref{fig:Figure4}(b), the wave propagates in the $y$-direction predominantly reflecting the collimation effect~\cite{25}. 

To investigate further, we perform the fast Fourier transformation (FFT) on the displacement field in the two spatial directions and plot the spectrum in Figs.~\ref{fig:Figure4}(c) and \ref{fig:Figure4}(d). We observe the arc-like pattern of the peak spectral density (in yellow). These are called ``Fermi arcs", also seen as the counterpart representation of the surface states in the reciprocal space. Since the normal vector to the Fermi arcs would determine the direction of the wave's group velocity, we can deduce from Fig.~\ref{fig:Figure4}(d) that the wave will propagate in the $k_y$-direction predominantly given the straight posture of the arc. This thus confirms the aforementioned collimation effect [Fig.~\ref{fig:Figure4}(b)] in the wavevector space.

These Fermi arcs could also be obtained from the equi-frequency contour (FEC) analysis on the supercell (see details in Appendix C).
To achieve this, we calculate dispersion characteristics of the supercells, as shown in Figs.~\ref{fig:Figure3}(b) and \ref{fig:Figure3}(c), for \textit{all} values of $k_z$ inside the first Brillouin zone. We then extract the wave numbers for $5.4$ kHz to obtain the red curves in Figs.~\ref{fig:Figure4}(c) and \ref{fig:Figure4}(d). Here, solid curves correspond to the the solutions on the plane of excitation, while the dashed curves represent those on the opposite surface. Evidently, the red solid curves closely match the spectral density arcs obtained from the full-scale simulation. The Fermi arc generally connects the Weyl points of the opposite charges~\citep{23}, but here we see that they connect the two Weyl points (in purple and yellow) roughly but not exactly. This is because the system supports the two Weyl points at different frequencies [see the frequency offset of the yellow and purple points in Fig.~\ref{fig:Figure2}(d)].

\begin{figure}[t]
\centering
\includegraphics[width=3.4in]{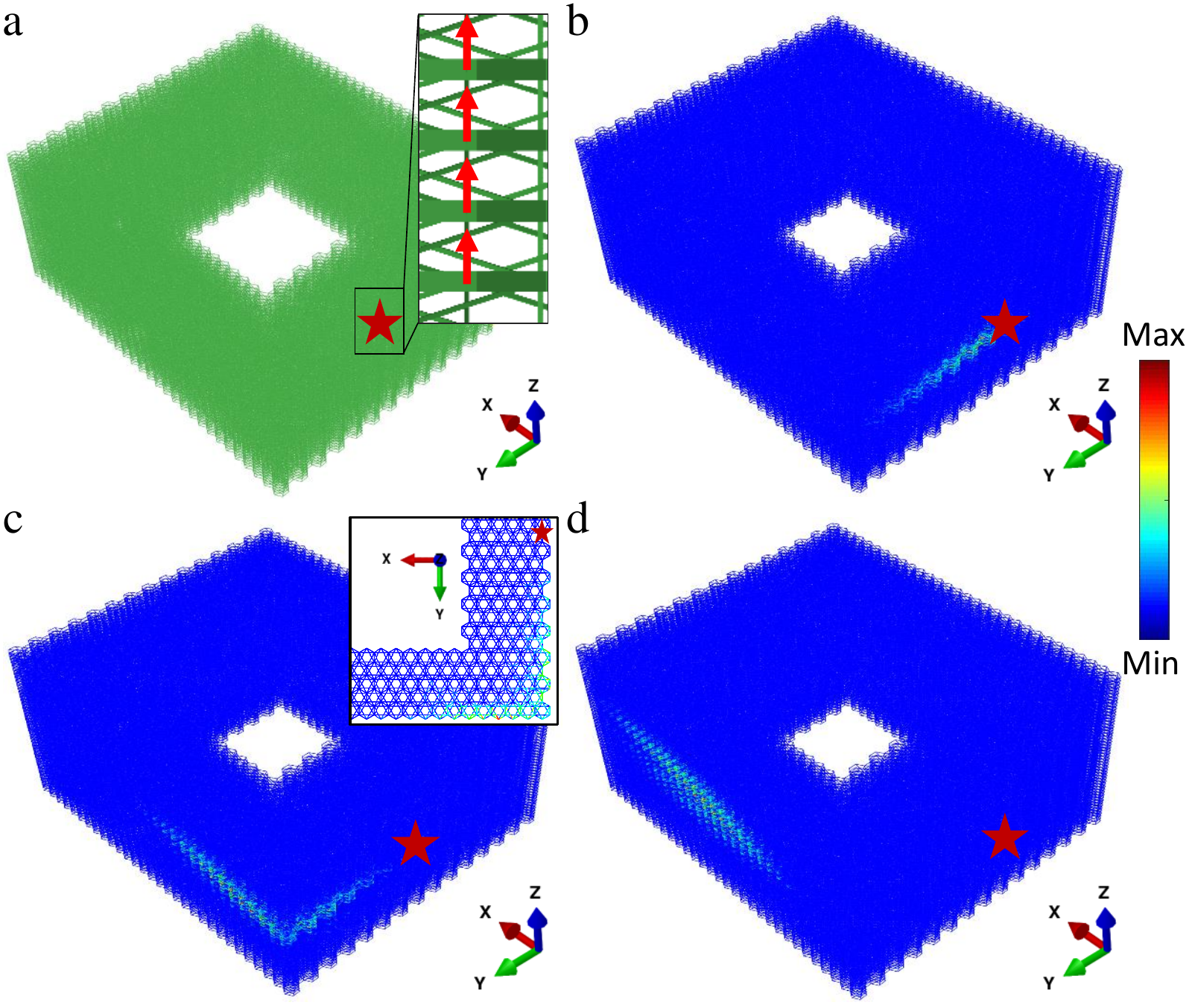}
\caption{[Color online] Robust one-way surface state propagating in a 3D hollow structure. (a) Star highlights the location of multi-point phased excitation. The exact locations of external loads are shown in the inset. (b)--(d) The magnitude of total displacement at time $t=4$ ms, $t=6.5$ ms, and $t=9$ ms, respectively. Inset in (c) shows the top view when the wave turns at the corner.}
\label{fig:Figure5}
\end{figure}

\subsection*{Robust one-way propagation}
We now proceed to the full-scale, transient numerical study performed at 5.4 kHz, but for a fixed $k_z$. Figure \ref{fig:Figure5}(a) shows the entire structure. We fix all the degrees of freedom of the nodes on the top and bottom layers. To ensure the excitation with the desired $k_z$, we apply four-point excitation in the $z$-direction on the $10$th to the $13$th layers on the YZ plane (marked with the red arrows in Fig. \ref{fig:Figure5}(a) inset).
We use a Gaussian-modulated sinusoidal pulse with the center frequency of $5.4$ kHz, and we fix $k_z=-\pi/(2P)$ by increasing the phase of the input signal by $\pi/2$ from the $10$th to the $13$th layer.  From the discussions above, we expect that the wave packet would propagate \textit{clockwise} when looking from the top ($z$-axis).
In Figs. \ref{fig:Figure5}(b)--\ref{fig:Figure5}(d), we plot the displacement amplitude of the system at time $t=4$ ms, $t=6.5$ ms, and $t=9$ ms, respectively. We observe that the elastic wave remains on the surface of the structure and only travels upward in the clockwise direction (viewing from the top) without obvious reflection at the corners  (see Fig. \ref{fig:Figure5}(c) and its inset, also Supplementary Movie 1). This, therefore, demonstrates a robust one-way propagation of surface elastic wave in our Weyl structure.

\section*{CONCLUSION}
We design a 3D mechanical structure -- analogous to the AA-stacked honeycomb lattice -- by using slender beams. We show that this relatively simple design carries Weyl points at the vertices of the Brillouin zone. We use a two-band Hamiltonian model to describe the dynamics around the Weyl points and calculate their topological charges.
We show the finite boundaries of this structure, both zigzag and armchair types, host localized states at a fixed $k_z$. 
Using numerical simulations on a full-scale 3D structure, we show the existence of the Fermi arcs and compare them with the results obtained from equi-frequency contour analysis. We highlight two unique wave phenomena in our structure: (i) collimation of the propagating elastic waves and (ii) robust one-way transport of elastic energy around the corners. 
Our design could be easily scaled up or down, and can be relevant to applications, such as sensing, energy harvesting, and vibration control on 3D elastic  structures. Studies on the experimental verification of the elastic surface states are expected in the authors' future publications.

\section*{ACKNOWLEDGMENTS}
We thank Dr. Hyunryung Kim, Dr. Ying Wu, and Shuaifeng Li for fruitful discussions. X. S., R. C., and J. Y. are grateful for the financial support from the U.S. National Science Foundation (CAREER1553202 and EFRI-1741685).

\section*{APPENDIX A: TIGHT-BINDING MODEL OF AA-STACKED GRAPHENE}
We consider the tight-binding model of a AA-stacked graphene with chiral interlayer coupling, as depicted in Fig. {\ref{fig:Figure1}}(a). The unit cell has an in-plane lattice constant $L$ and out-of-plane lattice constant $P$ in the $z$-direction [see Fig. {\ref{fig:Figure1}}(b)]. Let the intralayer (interlayer) coupling be $t_n$ ($t_c$). Therefore, we write the Bloch Hamiltonian given by~\cite{16,21,27} 

$$ H(\textbf{k}) = \left( {\begin{array}{*{20}{c}} {{\varepsilon}+{t_c}f({k_z}P)} & {{t_n}\beta }\\ { \left({t_n}\beta \right)^*}& {{\varepsilon}+{t_c}f( - {k_z}P)} \end{array}} \right), $$
where ${\varepsilon}$ denotes the on-site potential, $\beta=e^{-ik_yL}+2 e^{ik_yL/2} \cos \left(\sqrt{3}k_xL/{2} \right) $, and $f(k_zP)=2 \cos(\sqrt{3} k_x L-k_zP)+4 \cos(3k_yL/{2}) \cos(\sqrt{3}k_xL/{2}+k_zP)$.
By applying the $k \cdot p$ method\cite{16}, we can expand the Hamiltonian near the $K$ point [$k_x={4\sqrt{3}\pi}/(9L),k_y=0,k_z=0$] and obtain the effective Hamiltonian
\begin{eqnarray}
\begin{aligned}
H(\D \textbf{k}) =& (\varepsilon-3t_c)\sigma_0-\frac{3}{2}Lt_n(\D k_x\sigma_x-\D k_y\sigma_y) \\
&- 3\sqrt{3}Lt_c\D k_z\sigma_z, \nonumber
\end{aligned}
\end{eqnarray}
where $\D \textbf{k}=(\D k_x, \D k_y,\D k_z)$ is a small $k$-vector deviating from the $K$ point, $\sigma_0$ is the $2 \times 2$ unit matrix, and $\sigma_x$, $\sigma_y$, $\sigma_z$ are the Pauli matrices. 

We use $\varepsilon=5.73$ kHz, $t_n=0.875$ kHz, and $t_c=-0.131$ kHz for fitting the two-band dispersion with the curves obtained from the FEA results seen in Fig. {\ref{fig:Figure2}}.
The effective Hamiltonian describes a Weyl point at the $K$ point, whose topological charge is given by $C=\sgn(v_xv_yv_z)$, where Dirac velocities $v_x=-3L t_n/2$, $v_y=3 L t_n/2$, and $v_z=-3\sqrt{3}L t_c$. Therefore, $C=-1$ in this case. Similarly, by expanding the Hamiltonian at the  $H$ point [$k_x={4\sqrt{3}\pi}/(9L), k_y=0,k_z=\pi/P$], we verify that there is another Weyl point with the topological charge of $+1$ located at the $H$ point.

\section*{APPENDIX B: UNIT-CELL DISPERSION AND MODE POLARIZATION}
\begin{figure}[b!]
\centering
\includegraphics[width=3.4in]{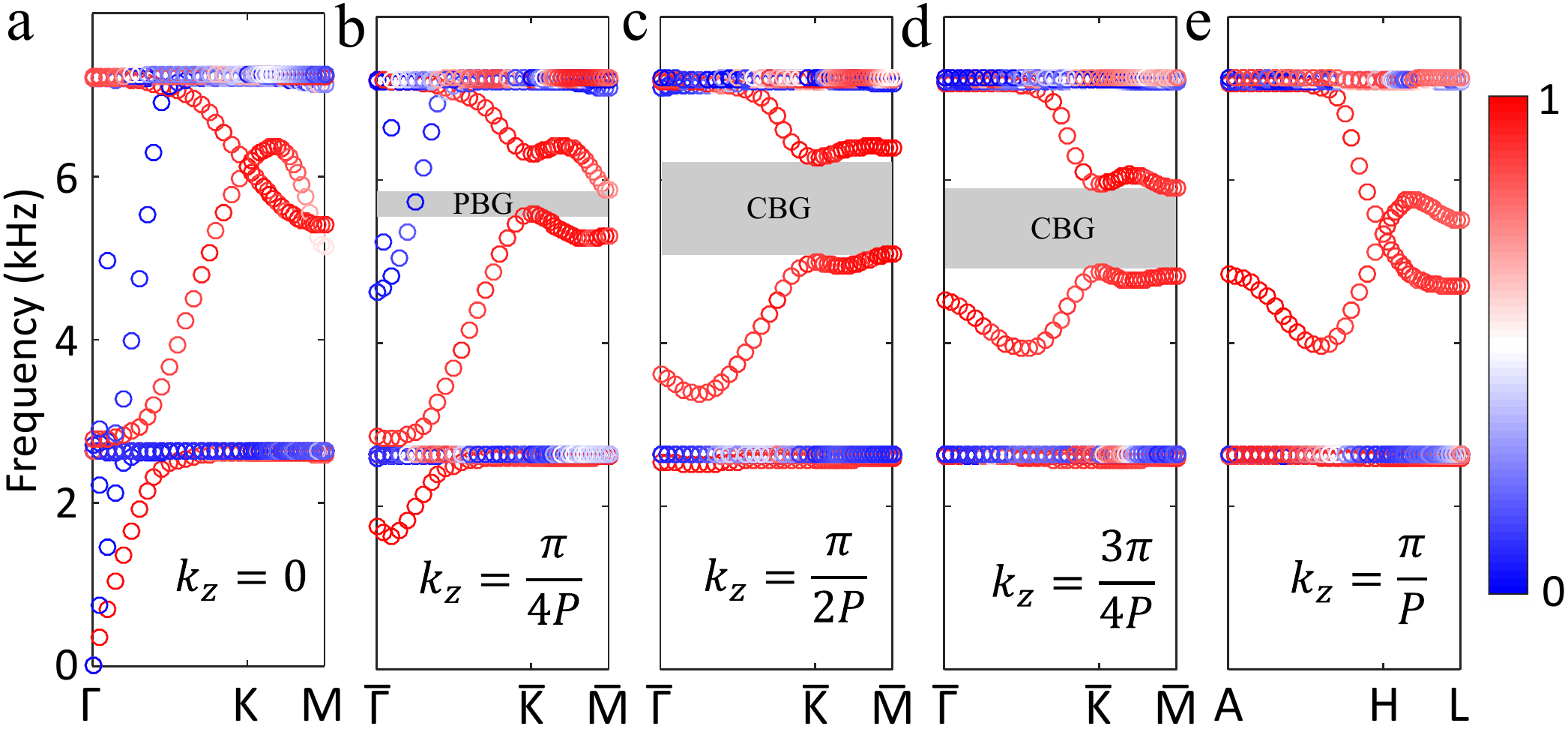}
\caption{[Color online] Dispersion diagram of the unit cell in the $k_xk_y$ plane at (a) $k_z=0$, (b) $k_z=\pi/(4P)$, (c) $k_z=\pi/(2P)$, (d) $k_z=3\pi/(4P)$, and (e) $k_z=\pi/P$, respectively. Colormap represents the polarization factor $P_z$. The grey area in (b) represents a \textit{partial} frequency band gap (noted as PBG) while the grey zones in (c) and (d) refer to \textit{complete} band gaps (noted as CBG). The flat branches mainly represent the cases when the interlayer beams are locally resonant. }
\label{fig:Figure6}
\end{figure}

In this appendix, we show the modes that are degenerate at the Weyl points are an \textit{out-of-plane}  type with a predominant $z$ component. To this end, we define a polarization factor $P_z= {|U_z|^2}/({|U_x|^2+|U_y|^2+|U_z|^2}) $ to distinguish bands with different polarization components, where $U_x$, $U_y$, and $U_z$ are the $x$-, $y$-, and $z$- components of the eigenvectors. Therefore, the out-of-plane modes, with predominately $U_z$ component, would have the polarization factor close to a unity, while the polarization factors of the in-plane modes would approach zero. We plot the bulk bands of the unit cell -- colored with the information about the polarization factors -- on the 2D reciprocal plane at various $k_z$ values in Fig. {\ref{fig:Figure6}}. We can clearly see that the Weyl points are formed by the degeneracy of the two bands containing out-of-plane modes (in red). 

\section*{APPENDIX C: EQUI-FREQUENCY CONTOUR ANALYSIS}

\begin{figure}[t]
\centering
\includegraphics[width=3.4in]{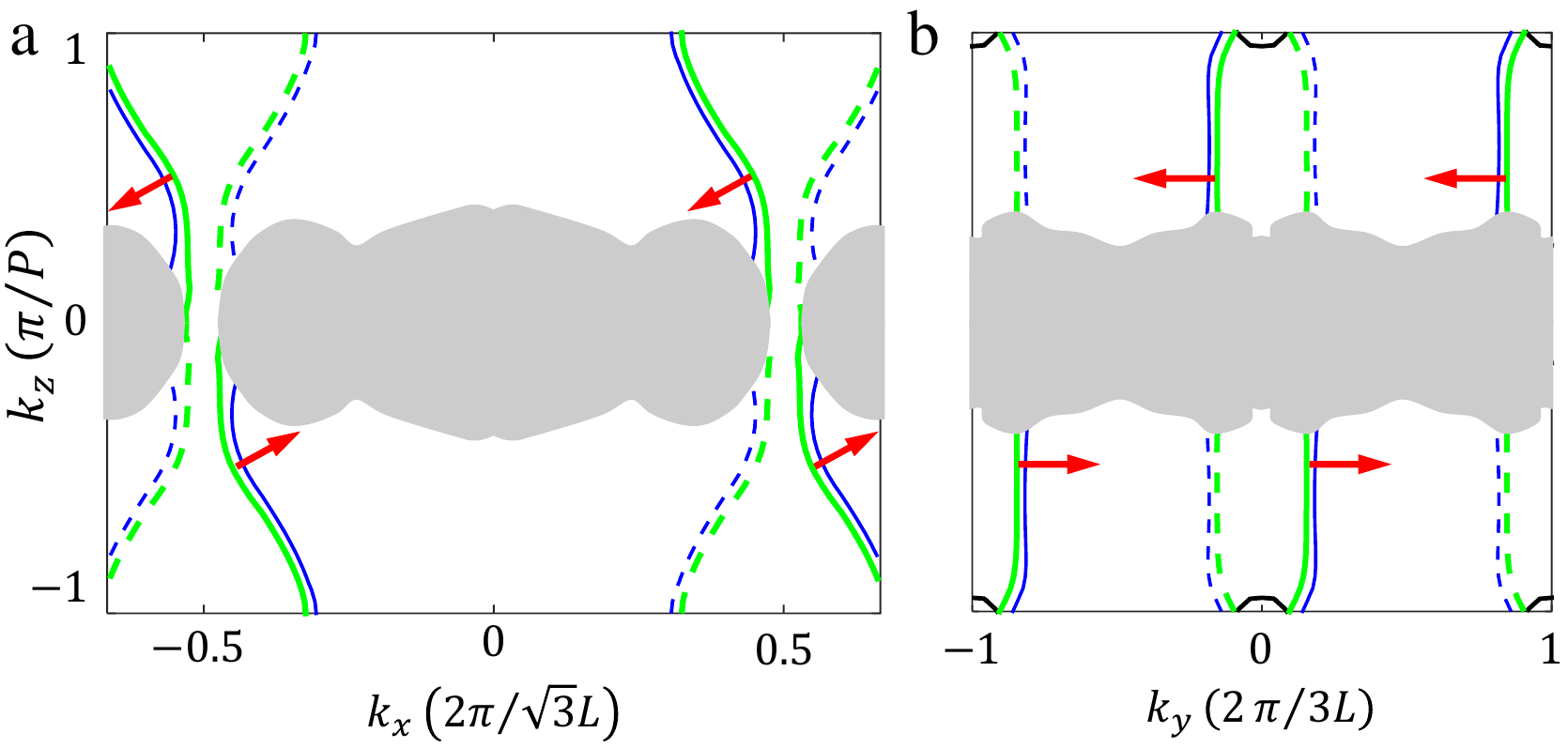}
\caption{[Color online] (a) Equi-frequency contour of the supercell with the zigzag type boundaries at $f=5.4$ kHz (green) and $f=5.5$ kHz (blue). Red arrows show the normal vectors of the dispersion curves. Solid and dashed lines represents the mode in the forward and the opposite planes as indicated in Fig. {\ref{fig:Figure4}}. The grey regions refers to the projections of the bulk bands. (b) Similar results of the supercell with armchair type boundaries.}
\label{fig:Figure7}
\end{figure}

Equi-frequency contour on the the $xz$ plane, as shown in Fig. {\ref{fig:Figure4}}(c), can be obtained by calculating the band structure of the supercell with the zigzag type boundary. At $f=5.4$ kHz, we vary $k_x$ from $-4\pi/(3\sqrt3L)$ to $4\pi/(3\sqrt3L)$ and $k_z$ from $-\pi/P$ to $\pi/P$, to obtain 
wavevector plots in green as shown in Fig. {\ref{fig:Figure7}}(a). Note that we plot another equi-frequency plot at slightly higher frequency ($f=5.5$ kHz) in blue to get a sense of the normal vector, i.e., the group velocity shown by red arrows at $k_z = \pm \pi/2P$. Similarly, result of the $yz$ plane, from the supercell with the armchair type boundary, is shown in Fig. {\ref{fig:Figure7}}(b). This equi-frequency contour confirms the collimation effect in the $k_y$ direction.


\begin{thebibliography}{99}

\bibliographystyle{plain}
\bibitem{1} Z. Liu, X. Zhang, Y. Mao, Y. Y. Zhu, Z. Yang, C. T. Chan, and P. Sheng, Locally Resonant Sonic Materials, \textit{Science} \textbf{289}, 1734 (2000).
\bibitem{2} M. I. Hussein, M. J. Leamy, and M. Ruzzene, Dynamics of Phononic Materials and Structures: Historical Origins, Recent Progress, and Future Outlook, \textit{Appl. Mech. Rev.} \textbf{66}, 040802 (2014).
\bibitem{3} M. Kadic, T. Bückmann, R. Schittny, and M. Wegener, Metamaterials beyond electromagnetism, \textit{Rep. Prog. Phys.} \textbf{76}, 126501 (2013).
\bibitem{4} K. Bertoldi, V. Vitelli, J. Christensen, and M. van Hecke, Flexible mechanical metamaterials, \textit{Nature Reviews Materials} \textbf{2}, 17066 (2017).

\bibitem{5} M. Z. Hasan and C. L. Kane, Colloquium: Topological insulators, \textit{Rev. Mod. Phys.} \textbf{82}, 3045 (2010).
\bibitem{6} X.-L. Qi and S.-C. Zhang, Topological insulators and superconductors, \textit{Rev. Mod. Phys.} \textbf{83}, 1057 (2011).
\bibitem{7} E. Prodan and H. Schulz-Baldes, Bulk and Boundary Invariants for Complex Topological Insulators (Springer International Publishing, Cham, 2016) arXiv:1510.08744.

\bibitem{8} S. D. Huber, Topological mechanics, \textit{Nature Physics} \textbf{12}, 621 (2016).
\bibitem{9} G. Ma, M. Xiao, and C. T. Chan, Topological phases in acoustic and mechanical systems, \textit{Nature Reviews Physics} \textbf{1} (2019).


\bibitem{10} X. Wan, A. M. Turner, A. Vishwanath, and S. Y. Savrasov, Topological semimetal and Fermi-arc surface states in the electronic structure of pyrochlore iridates, \textit{Phys. Rev. B} \textbf{83}, 205101 (2011).
\bibitem{11} C. Fang, M. J. Gilbert, X. Dai, and B. A. Bernevig, Multi-Weyl Topological Semimetals Stabilized by Point Group Symmetry, \textit{Phys. Rev. Lett.} \textbf{108}, 266802 (2012).
\bibitem{12} S.-Y. Xu, I. Belopolski, N. Alidoust, M. Neupane, G. Bian, C. Zhang, R. Sankar, G. Chang, Z. Yuan, C.-C. Lee, S.-M. Huang, H. Zheng, J. Ma, D. S. Sanchez, B. Wang, A. Bansil, F. Chou, P. P. Shibayev, H. Lin, S. Jia, and M. Z. Hasan, Discovery of a Weyl fermion semimetal and topological Fermi arcs, \textit{Science} \textbf{349}, 613 (2015).
\bibitem{13}A. A. Soluyanov, D. Gresch, Z. Wang, Q. Wu, M. Troyer, X. Dai, and B. A. Bernevig, Type-II Weyl semimetals, \textit{Nature} \textbf{527}, 495 (2015).
\bibitem{14} N. P. Armitage, E. J. Mele, and A. Vishwanath, Weyl and Dirac semimetals in three-dimensional solids, \textit{Rev. Mod. Phys.} \textbf{90}, 015001 (2018).




\bibitem{15} Z. Fang, N. Nagaosa, K. S. Takahashi, A. Asamitsu, R. Mathieu, T. Ogasawara, H. Yamada, M. Kawasaki, Y. Tokura, and K. Terakura, The Anomalous Hall Effect and Magnetic Monopoles in Momentum Space, \textit{Science} \textbf{302}, 92 (2003).

\bibitem{16} W.-J. Chen, M. Xiao, and C. T. Chan, Photonic crystals possessing multiple Weyl points and the experimental observation of robust surface states, \textit{Nature Communications} \textbf{7}, 13038 (2016).

\bibitem{17} H. B. Nielsen and M. Ninomiya, The Adler-Bell-Jackiw anomaly and Weyl fermions in a crystal, \textit{Physics Letters B} \textbf{130}, 389 (1983).

\bibitem{18} L. Lu, L. Fu, J. D. Joannopoulos, and M. Soljačić, Weyl points and line nodes in gyroid photonic crystals, \textit{Nature Photonics} \textbf{7}, 294 (2013).
\bibitem{19} L. Lu, Z. Wang, D. Ye, L. Ran, L. Fu, J. D. Joannopoulos, and M. Soljačić, Experimental observation of Weyl points, \textit{Science} \textbf{349}, 622 (2015).
\bibitem{20} M.-L. Chang, M. Xiao, W.-J. Chen, and C. T. Chan, Multiple Weyl points and the sign change of their topological charges in woodpile photonic crystals, \textit{Phys. Rev. B} \textbf{95}, 125136 (2017).

\bibitem{21} M. Xiao, W.-J. Chen, W.-Y. He, and C. T. Chan, Synthetic gauge flux and Weyl points in acoustic systems, \textit{Nature Physics} \textbf{11}, 920 (2015).
\bibitem{22} Z. Yang and B. Zhang, Acoustic Type-II Weyl Nodes from Stacking Dimerized Chains, \textit{Phys. Rev. Lett.} \textbf{117}, 224301 (2016).
\bibitem{23} F. Li, X. Huang, J. Lu, J. Ma, and Z. Liu, Weyl points and Fermi arcs in a chiral phononic crystal, \textit{Nature Physics} \textbf{14}, 30 (2018).
\bibitem{24} T. Liu, S. Zheng, H. Dai, D. Yu, and B. Xia, Acoustic semimetal with Weyl points and surface states, arXiv:1803.04284 [Cond-Mat] (2018).
\bibitem{25} H. Ge, X. Ni, Y. Tian, S. K. Gupta, M.-H. Lu, X. Lin, W.-D. Huang, C. T. Chan, and Y.-F. Chen, Experimental Observation of Acoustic Weyl Points and Topological Surface States. \textit{Phys. Rev. Applied} \textbf{10}, 014017 (2018).
\bibitem{26} H. He, C. Qiu, L. Ye, X. Cai, X. Fan, M. Ke, F. Zhang, and Z. Liu, Topological negative refraction of surface acoustic waves in a Weyl phononic crystal, \textit{Nature} \textbf{560}, 61 (2018).
\bibitem{27} X. Zhang, M. Xiao, Y. Cheng, M.-H. Lu, and J. Christensen, Communications Physics 1, 97 (2018).

\bibitem{28} M. Fruchart, S.-Y. Jeon, K. Hur, V. Cheianov, U. Wiesner, and V. Vitelli, Soft self-assembly of Weyl materials for light and sound, \textit{PNAS} \textbf{115}, E3655 (2018).
\bibitem{29} Y.-T. Wang and Y.-W. Tsai, Multiple Weyl and double-Weyl points in an elastic chiral lattice, \textit{New J. Phys.} \textbf{20}, 083031 (2018).

\bibitem{30} P. Wang, F. Casadei, S. H. Kang, and K. Bertoldi, Locally resonant band gaps in periodic beam lattices by tuning connectivity, \textit{Phys. Rev. B} \textbf{91}, 020103 (2015).

\bibitem{31} C. S. Lefky, B. Zucker, D. Wright, A. R. Nassar, T. W. Simpson, and O. J. Hildreth, Dissolvable Supports in Powder Bed Fusion-Printed Stainless Steel,\textit{3D Printing and Additive Manufacturing} \textbf{4}, 3 (2017).

\bibitem{32} F. D. M. Haldane, Model for a quantum Hall effect without Landau levels: Condensed-matter realization of the "Parity Anomaly", \textit{Phys. Rev. Lett.} \textbf{61}, 2015 (1988).




\end{thebibliography}
\end{document}